\documentclass[10pt,twocolumn,twoside]{IEEEtran}
\usepackage{url}
\IEEEoverridecommandlockouts

\usepackage{cite}
\usepackage{amsmath,amssymb,amsfonts}
\usepackage{algorithmic}
\usepackage{graphicx}
\usepackage{textcomp}
\usepackage{balance}
\usepackage{xcolor}
\usepackage[utf8]{inputenc}
\usepackage{multirow}
\usepackage[nolist]{acronym}
\usepackage[final]{changes}

\def\BibTeX{{\rm B\kern-.05em{\sc i\kern-.025em b}\kern-.08em
    T\kern-.1667em\lower.7ex\hbox{E}\kern-.125emX}}
\usepackage{lipsum}

\title{\LARGE \bf 
On the Concept of an Optimal Portfolio of Uncertain Flexible Loads
}

\author{\added{Julie Rousseau$^{*,\dagger}$, Philipp Heer$^\dagger$, Kristina Orehounig$^\ddagger$, Gabriela Hug$^*$}
\vspace{-0.5cm}
 \thanks{ \added{This work was supported as a part of NCCR Automation, a National Centre of Competence in Research, funded by the Swiss National Science Foundation (grant number 51NF40\_225155).}}
 \thanks{\added{$^*$ Julie Rousseau and Gabriela Hug are with the Power Systems Laboratory, ETH Zürich, Switzerland. Email: {\tt \{jrousseau, ghug\}@ethz.ch}.}}
\thanks{\added{$^\dagger$Julie Rousseau and Philipp Heer are with the Urban Energy Systems Laboratory, Empa, Dübendorf, Switzerland. }}
\thanks{\added{$^\ddagger$Kristina Orehounig is with the Research Unit for Building Physics and Building
Ecology, Vienna University of Technology, Vienna, Austria.}}}

\begin{document}
\begingroup
\allowdisplaybreaks

\maketitle

\begin{abstract}
Flexible loads can enhance power system stability by providing reserves, but their limited energy capacity and uncertain availability distinguish them from conventional generators. To accommodate these characteristics, the Danish Transmission System Operator (TSO) recently introduced new reserve market rules that incorporate energy constraints and relax reliability requirements. In this context, the optimal reserve quantification becomes a joint chance-constrained reserve quantification problem, which is difficult to solve. In this paper, we derive two analytical reformulations of this problem: an exact one when a reserve direction dominates and an approximate one, otherwise. Furthermore, when flexible loads must collectively satisfy a reliability requirement, we introduce the concept of an optimal portfolio of flexible loads: adding loads with similar expected values but different stochastic behaviors to an existing portfolio may change the total portfolio's reserves. To support this idea, we theoretically study the marginal increase in reserves resulting from adding a load. Numerical results show that our analytical reformulations closely match the exact formulation, with a mean absolute error of 2.5\%. Case studies further demonstrate the existence of optimal load groupings and our ability to predict the portfolio in which a load’s marginal value is highest, leveraging our theoretical analysis.
\end{abstract}


\begin{acronym}[ML] 
    \acro{TSO}{Transmission System Operator}
    \acro{EV}{Electric Vehicle}
    \acro{HVAC}{Heating, Ventilation and Air Conditioning}
\end{acronym}

\section{Introduction}
In an attempt to reduce global carbon emissions, electric power grids undergo profound transformations: a large number of flexible carbon-intensive power plants are decommissioned and replaced by intermittent renewable energy sources \cite{irena2019}. While conventional power plants used to constantly adapt their production to the electric power demand, renewable power units can only partly do so. 
Hence, policymakers encourage power consumers to become more flexible, i.e., partly adapt their consumption by optimally planning it in advance or providing ancillary services in real time  \cite{EUFlexibility}.


In the residential sector, \added{Heating, Ventilation, and Air Conditioning (HVAC) systems} and Electric Vehicles (EVs) are among the most promising sources of flexibility \cite{irena2019}. The power consumption of an HVAC can be shifted over time, as long as the rooms it serves remain within predefined temperature bounds. Similarly, the power consumed by an EV can be adjusted, as long as the user's requested state-of-charge at departure is guaranteed. Additionally, the power consumption of both devices is also limited by their power and energy technical ratings. Previous research has proven the technical viability of ancillary service provision by such flexible loads \cite{vrettos2018, BUNNING2022}.

Compared to conventional generators, flexible loads are limited both in the amount of power and energy they can consume over time \cite{PLAUM2022}. 
Yet, most ancillary service products are specified in terms of ramping rates, power, and expected duration \cite{RANCILIO2022}, but do not specify the amount of energy to be reserved. A crude approach consists in considering the worst-case energy activation, i.e., the delivery of the full power over the duration \cite{rousseau2026}. While robust, this approach also appears overly conservative, especially for durations of multiple hours or days. Instead, an active energy management is preferable. It consists of continuously adjusting the loads' energy contents through intra-day trades to offset previous energy activations \cite{koller2016, BUNNING2022, rousseau2026, GORECKI2017229}. 

Furthermore, the flexibility of HVACs and EVs is also partly uncertain and, therefore, complex to plan. Indeed, an EV's flexibility strongly depends on when the vehicle leaves, which, in most cases, is uncertain \cite{sun2020}. 
The flexibility of an HVAC is also hard to predict, as it depends on uncertain future weather conditions, the presence and actions of inhabitants, and thermal models with limited accuracy \cite{rousseau2026}. 
However, ancillary service markets require the offered flexibility to be almost certain, placing the burden on aggregators to manage risk, e.g., by trading off revenues for flexibility with penalties for failing to deliver the promised reserves \cite{rousseau2026, Gade2023}. 

Acknowledging the current lack of specifications for energy and reliability in ancillary service markets, new market rules have been developed in Denmark \cite{Gade2024} to attract more flexible loads. Specifically, it defines an amount of energy to be reserved by energy-constrained loads, as well as an amount of power to be reserved in the opposite direction \cite{energinet}. Both specifications give formal requirements for the implementation of an energy management system in practice. Additionally, the new market rules specify a reduced reliability level, requiring loads to be available with a 90\% probability, down from the previous 99.9\% \cite{energinet}. This reliability requirement shifts part of the burden of managing risk from aggregators to TSOs.

Given these new energy and reliability specifications, some studies explore how flexible loads should participate in the Danish reserve markets. The problem is formulated as a joint-chance constrained optimization, which is generally complex to solve. Existing studies solve the optimization problem using various methods, e.g, the iterative ALSO-X method \cite{Lunde2025}, the conservative CVaR method \cite{Lunde2025}, or a distributionally robust reformulation \cite{Gade2024}. Alternatively, the authors of \cite{Herstad2026} suggest separating the joint chance constraints into individual chance constraints using the Bonferroni approximation. While all methods are valid, the authors of \cite{zapparoli2025} observe that an analytical reformulation of the problem exists, offering a computationally efficient, non-conservative approach. However, this approach assumes that loads can only reduce their consumption and no comparison with existing methods is provided. Following a similar approach to \cite{zapparoli2025}, we derive an analytical formulation to determine the reserves of a portfolio of loads in both directions and compare it to existing methods.

Besides, when grouping loads, previous studies observe that larger portfolios can provide greater reserves relative to their number of loads \cite{Lunde2025, Gade2023}. Another phenomenon likely affects a portfolio's reserves: for a given number of loads, grouping different stochastic profiles can yield different reserve amounts. In this paper, we explore the concept of an optimal portfolio, i.e., how to group loads based on their stochastic behavior to maximize the portfolio's reserve amount. The contribution is, therefore, three-fold: 
\begin{itemize}
    \item We propose an analytical method to determine the maximum amount of reserves a portfolio can offer under the new Danish rules. Technically, we propose an analytical reformulation of a joint chance-constrained optimization.
    \item \added{We demonstrate a counter-intuitive result showing how wisely pairing loads into different portfolios} may increase the total amount of reserves for a given reliability level.
    \item We develop theoretical foundations to understand the marginal increase in reserves, resulting from adding a load to a portfolio. Specifically, we study the impact of certain load and portfolio characteristics.
\end{itemize}

The remainder of this paper is organized as follows. Section~\ref{sec:methodology} presents an analytical formulation to determine the optimal participation of a portfolio, i.e., a set of flexible loads, in the Danish reserve markets. Section~\ref{sec:optimal_portfolio} further investigates the concept of an optimal portfolio, i.e., an optimal grouping of loads according to their stochastic properties. Then, Section~\ref{sec:case_study} describes the case study, Section~\ref{sec:results} analyzes the results, and Section~\ref{sec:conclusion} concludes the work. 

In the remainder of this paper, the sets $\mathcal{T}$ and  $\mathcal{I}$ describe the discrete set of timesteps and loads, respectively. Tildes denote stochastic variables, and $\Delta t$ designates the time resolution between two discrete timesteps $k$ and $k+1$.

\section{A Portfolio of Flexible Loads Participating in the Danish Reserve Markets}
\label{sec:methodology}

This section first describes the flexibility potential of two types of loads, EVs and HVACs, before explaining how they can sell their flexibility in the Danish reserve markets.

\subsection{Flexible Loads}
\label{subsec:flexible_loads}

\added{In this paper, we describe load flexibility through the concept of power and energy envelopes \cite{rousseau2025}, and limit ourselves to EVs and HVACs. The methodology extends naturally to any asset with limited power and energy consumption. }

\subsubsection{EVs}
Once plugged in, an electric vehicle is flexible in the sense that its power consumption can be shifted over time, as long as its final state-of-charge requirement is guaranteed. The EV $i$ arrives at a charging station at $t^a_{\mathrm{ev},i}$ with an initial energy content $e^0_{\mathrm{ev},i}$ and leaves at $t^d_{\mathrm{ev},i}$. Its power and energy consumption are limited by its technical ratings:
\begin{equation}
\begin{aligned}
    p_{\mathrm{ev},i}^{\mathrm{min},k} & \leq p_{\mathrm{ev},i}^k \leq p_{\mathrm{ev},i}^{\mathrm{max},k}, \hspace{2.8cm} \forall k \in \mathcal{T}, \\
    e_{\mathrm{ev},i}^{\mathrm{min},k} & \leq e^0_{\mathrm{ev},i} + \Delta t \sum_{t = t^a_{\mathrm{ev},i}}^k p_{\mathrm{ev},i}^t \leq e_{\mathrm{ev},i}^{\mathrm{max},k}, \quad \forall k \in \mathcal{T},
\end{aligned}
\label{eq:power_energy_bounds_ev}
\end{equation}
where $p_{\mathrm{ev},i}^{\mathrm{min},k}$,  $p_{\mathrm{ev},i}^{\mathrm{max},k}$, $e_{\mathrm{ev},i}^{\mathrm{min},k}$, and $e_{\mathrm{ev},i}^{\mathrm{max},k}$ describe its time-varying minimum and maximum power and energy ratings, respectively. \added{The power} ratings are equal to the respective EV rating when plugged-in, e.g., $p_{\mathrm{ev},i}^{\mathrm{min}}$ for $p_{\mathrm{ev},i}^{\mathrm{min},k}$, and 0 otherwise. \added{The time-varying energy bounds correspond to the minimum and maximum energy consumption paths, as described in \cite{Madjidian2018}.} Finally, upon departure, at $t^d_{\mathrm{ev},i}$, EV $i$ should be charged to at least the level of $e^f_{\mathrm{ev},i}$, i.e.:
\begin{equation}
    e^0_{\mathrm{ev},i} + \Delta t \sum_{t = t^a_{\mathrm{ev},i}}^k p_{\mathrm{ev},i}^t \geq e^f_{\mathrm{ev},i}, \quad \forall k \geq t^d_{\mathrm{ev},i}.
    \label{eq:energy_at_departure_ev}
\end{equation}
Constraint (\ref{eq:energy_at_departure_ev}) can be integrated into the time-varying lower energy bound $e_{\mathrm{ev},i}^{\mathrm{min},k}$ such that the flexibility of EV $i$ is simply delimited by time-varying power and energy bounds.

However, in practice, EVs' future flexibility is often uncertain, e.g., due to uncertain departure times. As a consequence, the time-varying power and energy bounds become stochastic variables denoted as $\tilde{p}_{\mathrm{ev},i}^{\mathrm{min},k}$,  $\tilde{p}_{\mathrm{ev},i}^{\mathrm{max},k}$, $\tilde{e}_{\mathrm{ev},i}^{\mathrm{min},k}$, and $\tilde{e}_{\mathrm{ev},i}^{\mathrm{max},k}$. 

\subsubsection{HVACs} Electric heating and cooling systems can adjust their power consumption while maintaining thermal comfort for inhabitants, often defined as an acceptable indoor temperature range. To relate the HVAC $i$'s power consumption $p_{\mathrm{hvac},i}^k$ to indoor temperatures, we rely on a discretized \added{first-order resistance-capacitance lumped model}:
\begin{equation}
    C_i\frac{T_i^{k+1} - T_i^k}{\Delta t} = -\frac{1}{R_i} \left(T_i^k - T_{a,i}^k\right) + p_{\mathrm{hvac},i}^k+ s_i^k,
\end{equation}
where $T_i^k$ describes the representative indoor temperature of the building heated/cooled by HVAC $i$, which initially equals $T_i^{0}$. Parameters $R_i$ and $C_i$ are the thermal resistance and capacitance of this building, respectively, whereas $T_{a,i}$ and $s_i$ describe the ambient air temperature and the solar heat gains, respectively. The power consumption of HVAC $i$ is limited by its technical power ratings, $p^{\mathrm{min}}_{\mathrm{hvac},i}$ and $p^{\mathrm{max}}_{\mathrm{hvac},i}$ as: 
\begin{equation}
    p^{\mathrm{min}}_{\mathrm{hvac},i} \leq p^k_{\mathrm{hvac},i} \leq p^{\mathrm{max}}_{\mathrm{hvac},i}, \quad \forall k \in \mathcal{T}.
    \label{eq:hvac_power_bounds}
\end{equation}
Additionally, to guarantee the inhabitants' thermal comfort, the indoor temperature must remain within an acceptable range: 
\begin{equation}
    T^{\mathrm{min}}_{i} \leq T^k_{i} \leq T^{\mathrm{max}}_{i}, \quad \forall k \in \mathcal{T}.
\end{equation}
Given these temperature constraints and following the methodology of \cite{rousseau2025}, we can determine the time-varying minimum and maximum energy HVAC $i$ can consume over a given horizon, denoted as $e^{\mathrm{min},k}_{\mathrm{hvac},i}$ and $e^{\mathrm{max},k}_{\mathrm{hvac},i}$, respectively. These time-varying energy bounds, together with the power bounds defined in (\ref{eq:hvac_power_bounds}), delimit the flexibility of  HVAC $i$. 

However, the time-varying energy bounds that define the future flexibility may be uncertain, e.g., due to uncertain future weather \cite{rousseau2026}. Hence, they become stochastic variables, denoted as $\tilde{e}^{\mathrm{min},k}_{\mathrm{hvac},i}$ and $\tilde{e}^{\mathrm{max},k}_{\mathrm{hvac},i}$. Additionally, to harmonize notations among loads, we also consider the power bounds to be time-varying stochastic variables, denoted as $\tilde{p}^{\mathrm{min},k}_{\mathrm{hvac},i}$ and $\tilde{p}^{\mathrm{max},k}_{\mathrm{hvac},i}$.

\subsection{The Danish FCR-D Market}

The Danish TSO recently introduced new requirements to enable uncertain energy-constrained flexibility providers to participate in ancillary service reserve markets \cite{energinet}. While these new rules apply to all Danish ancillary services, FCR-D has gained particular attention, since it is an hourly asymmetric product. The Frequency Containment Reserves for Disturbances (FCR-D) is an emergency reserve designed to support the regular FCR in case the frequency deviates from the nominal frequency by more than 0.1~Hz. Our methodology complies with the requirements of the FCR-D market, similarly to \cite{Lunde2025, Gade2024, zapparoli2025}, but could also be adapted to other Danish ancillary services using the requirements detailed in \cite{energinet}.

\subsubsection{Energy Requirement}
\label{sub-sub-sec:energy_req}
The Danish TSO proactively defines energy and power requirements for FCR-D reserves, which allow energy-constrained loads to participate in ancillary service markets by managing their energy content \cite{koller2016}. More specifically, the Danish TSO specifies that, for a given amount of reserves in one direction at an hour, the flexibility provider should be able to deliver the full power during one third of the hour, and reserve 20\% of the power in the opposite direction. From an aggregator's point of view, this new requirement can be integrated into an optimization problem, whose objective is to maximize, at each hour $h$, revenues as \added{\cite{Lunde2025}}: 
\begin{subequations}
    \begin{align}
        & \hspace{-0.5cm} \max_{p^+_{i,h}, p^-_{i,h}}  \quad  c^+_h \sum_{i\in \mathcal{I}} p^+_{i,h} + c^-_h \sum_{i\in \mathcal{I}} p^-_{i,h} \\
        \text{s.t. } \quad 
        & b^k_{i} + p_{i,h}^+ + 0.2 p_{i,h}^- \leq p^{\mathrm{max},k}_{i}, \hspace{0.33cm} \forall k \in \mathcal{T}_h, \forall i \in \mathcal{I}, \label{opt-cstr:det_1}\\
        & b_{i}^k - p_{i,h}^- - 0.2 p_{i,h}^+ \geq {p}^{\mathrm{min},k}_{i}, \hspace{0.4cm} \forall k \in \mathcal{T}_h, \forall i \in \mathcal{I},\label{opt-cstr:det_2} \\
        & e_{b,i}^k + \frac{1}{3} p^+_{i,h} \leq e_i^{\mathrm{max},k}, \hspace{1.33cm} \forall k \in \mathcal{T}_h, \forall i \in \mathcal{I},\label{opt-cstr:det_3} \\
        & e_{b,i}^k - \frac{1}{3} p^-_{i,h} \geq e_i^{\mathrm{min},k}, \hspace{1.4cm} \forall k \in \mathcal{T}_h, \forall i \in \mathcal{I}, \label{opt-cstr:det_4}\\
        & p^+_{i,h}, p^-_{i,h} \geq 0, \hspace{3.93cm} \forall i \in \mathcal{I},
    \end{align}
    \label{opt:deterministic}
\end{subequations}
where $\mathcal{T}_h$ designates the subset of $\mathcal{T}$ included in hour $h$, and $\left( p^{\mathrm{min},k}_{i}, p^{\mathrm{max},k}_{i}, e_i^{\mathrm{min},k}, e_i^{\mathrm{max},k} \right)$ are the time-varying bounds defined in Section~\ref{subsec:flexible_loads}. Parameters $b_{i}^k$ and $e_{b,i}^k$ are the baseline power and energy consumption of the load $i$, respectively. \added{$e_{b,i}^k$ corresponds to the accumulated baseline power consumption, i.e., $e_{b,i}^k = \Delta t \sum_{l=0}^k b_{i}^l$.} The total upward and downward\footnote{Throughout the paper, upward reserves indicate an increase in power consumption, equivalent to a decrease in production, while the downward reserves designate a decrease in consumption, equivalent to an increase in production.} reserves, given by $\sum_{i\in \mathcal{I}} p^+_{i,h}$ and $\sum_{i\in \mathcal{I}} p^-_{i,h}$ respectively, are compensated with the prices $c^+_h$ and $c^-_h$. The problem can be expressed in more compact form as: 
\begin{subequations}
    \begin{align}
        \max_{p^+_{i,h}, p^-_{i,h}} & \quad c^+_h \sum_{i\in \mathcal{I}} p^+_{i,h} + c^-_h \sum_{i\in \mathcal{I}} p^-_{i,h} \\
        \text{s.t. } \quad & p_{i,h}^+ + 0.2 p_{i,h}^- \leq p^{\mathrm{up},+}_{i,h}, \quad \forall i \in \mathcal{I}, \label{opt-cstr:det_vec_1}\\
         & p_{i,h}^- + 0.2 p_{i,h}^+ \leq p^{\mathrm{up},-}_{i,h}, \quad \forall i \in \mathcal{I}, \label{opt-cstr:det_vec_2}\\
         & 0 \leq p^+_{i,h} \leq e^{\mathrm{up},+}_{i,h}, \hspace{1.1cm}  \forall i \in \mathcal{I}, \label{opt-cstr:det_vec_3}\\
         & 0 \leq p^-_{i,h} \leq e^{\mathrm{up},-}_{i,h}, \hspace{1.1cm} \forall i \in \mathcal{I}, \label{opt-cstr:det_vec_4}
    \end{align}
    \label{opt:deterministic_vec}
\end{subequations}
where the parameters $\left( p^{\mathrm{up},+}_{i,h}, p^{\mathrm{up},-}_{i,h}, e^{\mathrm{up},+}_{i,h}, e^{\mathrm{up},-}_{i,h} \right)$ are derived from the parameters in (\ref{opt:deterministic})\footnote{\added{The parameter values for hour $h$ are defined as the the lowest parameter values among timesteps $k \in \mathcal{T}_h$.}}.
This problem can be separated into sub-problems for each flexible load of the portfolio. Each sub-problem can then be efficiently solved using the simplex method. 
As an example, Fig.~\ref{fig:simplex_example} illustrates the feasible space (gray area) defined by constraints (\ref{opt-cstr:det_vec_1})-(\ref{opt-cstr:det_vec_4}) for load $i$. This forms a convex space, composed of at most 6~corners. The optimal solution can be found by evaluating the objective function in all 6~corners.

\subsubsection{Reliability Requirement}
\label{sub-sub-sec:reliability_req}
\begin{figure}
    \vspace{-0.3cm}
    \hspace{0.6cm}
    \includegraphics[width=0.65\linewidth]{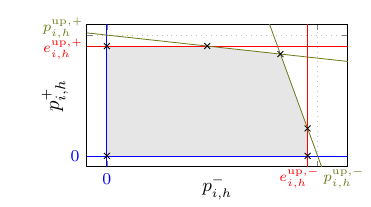}
    \vspace{-0.3cm}
    \caption{Feasible space delimited by constraints (\ref{opt-cstr:det_vec_1})-(\ref{opt-cstr:det_vec_4}) for load $i$.}
    \label{fig:simplex_example}
    \vspace{-0.5cm}
\end{figure} 

Additionally, the Danish TSO acknowledges the uncertainty associated with some flexible loads and allows them to provide reserves with a reduced reliability level of 90\%. In other words, whenever such flexible loads participate in ancillary service reserves, regardless if they are activated, their promised reserves should be available 90\% of the time. This additional requirement transforms problem (\ref{opt:deterministic_vec}) into the following joint chance-constrained optimization:
\begin{subequations}
    \begin{align}
        \max_{p^+_h, p^-_h} & \quad c^+_h  p^+_h + c^-_h p^-_h \\
        \text{s.t. } \quad & \mathbb{P}
        \begin{bmatrix}
            p_{i,h}^+ + 0.2 p_{i,h}^- \leq \tilde{p}^{\mathrm{up},+}_{i,h}, \quad \forall i \in \mathcal{I}, \\
            p_{i,h}^- + 0.2 p_{i,h}^+ \leq \tilde{p}^{\mathrm{up},-}_{i,h}, \quad \forall i \in \mathcal{I}, \\
            0\leq p^+_{i,h} \leq \tilde{e}^{\mathrm{up},+}_{i,h}, \hspace{1.1cm}\forall i \in \mathcal{I}, \\
            0\leq p^-_{i,h} \leq \tilde{e}^{\mathrm{up},-}_{i,h}, \hspace{1.1cm}\forall i \in \mathcal{I}, \\
            \hspace{-2.1cm} p^+_h \leq \sum_{i\in \mathcal{I}} p^+_{i,h}, \\
            \hspace{-2.1cm} p^-_h \leq \sum_{i\in \mathcal{I}} p^-_{i,h},
        \end{bmatrix} \geq R, \label{opt-cst:jjcc_1}
    \end{align}
    \label{opt:jjcc}
\end{subequations}
where $R = 90\%$ is the required reliability, and tilde variables are stochastic counterparts of the variables introduced in (\ref{opt:deterministic_vec}). By introducing $p^+_h$ and $p^-_h$, as sums of reserves from individual resources, we aim to provide a robust total amount of reserves rather than robust individual amounts. When a load fails to provide reserves, the flexibility of other loads can compensate for it, maintaining a high level of robust total reserves while the load's robust individual reserve level decreases. Hence, this reduces conservativeness.

To approach this problem, we first consider that one reserve direction is clearly more valuable\footnote{In this context, a direction is said to be clearly more valuable if the ratio of the smallest to the highest prices is lower than 0.2.} than the other, e.g., the upward direction. In this context, we first set the reserves in the opposite direction, i.e., $p^-_h$ in this case, to 0 such that constraint (\ref{opt-cst:jjcc_1}) can be formulated as:
\begin{equation}
    \mathbb{P}
        \begin{bmatrix}
            \hspace{0cm} p_{i,h}^+ \leq \tilde{p}^{\mathrm{up},+}_{i,h}, \hspace{1cm} \forall i \in \mathcal{I}, \\
            \hspace{0cm} 0.2 p_{i,h}^+ \leq \tilde{p}^{\mathrm{up},-}_{i,h}, \hspace{0.6cm} \forall i \in \mathcal{I}, \\
            0\leq p^+_{i,h} \leq \tilde{e}^{\mathrm{up},+}_{i,h}, \quad\forall i \in \mathcal{I}, \\
            \hspace{-1.7cm} p^+_h \leq \sum_{i\in \mathcal{I}} p^+_{i,h},
        \end{bmatrix} \geq R.
        \label{eq:stoc-constraint}
\end{equation}
The maximum of $p_h^+$ is obtained if the reserves provided by each load are maximized. In other words, each load reserves $p_{i,h}^+$ is equal to its upper bound, such that: 
\begin{equation}
    \mathbb{P} \left( p^+_h \leq \underbrace{\sum_{i \in \mathcal{I}} \min \left(  \tilde{p}^{\mathrm{up},+}_{i,h}, \frac{1}{0.2} \tilde{p}^{\mathrm{up},-}_{i,h}, \tilde{e}^{\mathrm{up},+}_{i,h} \right)}_{X_h^+},  \right) \geq R. 
\end{equation}
Hence, the optimal upward reserve amount $p^{+*}_h$ can be obtained analytically as the $(1-R)$ quantile of the stochastic variable\footnote{\added{This approach is similar to the one used in \cite{zapparoli2025}, although the authors do not explicitly define the random variable of interest, nor establish the validity of the approach.}} $ X_h^+$, i.e.:
\begin{equation}
    p^{+*}_h = q_{X_h^+} \left(1-R \right).
    \label{eq:quantile_decoupled}
\end{equation}
Such a quantile value can be evaluated using scenarios. \added{In this paper, we rely on a number of scenarios similar to \cite{Lunde2025}.}
Then, fixing the value of upward reserves to $p_h^{+*}$, we can determine the value of downward reserves using a similar technique\footnote{This approach is similar to the Naive approach developed in \cite{Lunde2025}. However, the authors of \cite{Lunde2025} considered the minimum of the three variables' quantile, instead of the quantile of the minimum of the three variables.}.

\added{When reserve prices are significantly different, reserves can be optimized sequentially. However, when prices are comparable, reserve values become coupled through the joint chance constraint, and the optimal reserve values no longer correspond to the quantiles of the marginal distributions. Therefore, applying the previous technique may yield suboptimal results.} In such a case, we suggest a scenario-based approach. For each scenario, we determine both total optimal reserves, using the vertex method to solve (\ref{opt:deterministic_vec}). Then, the optimization aims to find the values of $p^+_h$ and $p^-_h$ that maximize the objective function, while violating only $(1-R)$ scenarios. Therefore, we define the reserve values as the $(1-R)$ quantile of the joint distribution of the optimal values.
Nevertheless, it is important to note that this method is sub-optimal. Indeed, we first consider optimal pairs of values for each scenario before evaluating the joint quantile. Yet, in some scenarios, considering a sub-optimal couple may not change the overall quantile in one direction, but increase it in the opposite direction, leading to a higher final objective value.

\section{On the Concept of an Optimal Portfolio}
\label{sec:optimal_portfolio}

Given a portfolio of flexible loads, Section~\ref{sec:methodology} provides information about the optimal amount of reserves a portfolio can offer in the Danish reserve markets. However, this amount depends on its constituent loads. In this section, we investigate the design of a portfolio of flexible loads to maximize its reserve capabilities. 

Ignoring the reliability requirement introduced in Section~\ref{sub-sub-sec:reliability_req}, i.e., considering a deterministic case, a portfolio's reserves are equal to the sum of individual loads' reserves. Hence, for a fixed number of loads, the total amount of reserves across all portfolios remains identical regardless of how the loads are grouped into different portfolios. However, when considering a reliability requirement, this statement no longer holds true. Instead, some groupings may yield a higher total reserve amount, leading to the concept of optimal portfolios, i.e., an optimal grouping of a set of flexible loads to maximize the total reserves. \added{An illustrative example of this concept if provided in Appendix~\ref{app:example}.} In the following, we investigate the additional reserves a load brings to a portfolio.

\subsection{\added{Portfolio's Value Increase After Adding New Loads}}
\label{subsec:optimal_portfolio}

According to Section~\ref{sub-sub-sec:reliability_req}, the amount of reserves that a portfolio can offer relates to a specific quantile. Without loss of generality, for the rest of this section, we consider that one reserve direction is significantly more valuable, e.g., the upward direction. Therefore, the level of reserves that a portfolio can offer is:
\begin{equation}
    p^{+*}_h = q_{X_h^+} \left( 1-R \right) = q_{X} \left( 1-R \right),
\end{equation}
at hour $h$. For simplicity, here, we simply denote $X_h^+$ as $X$.


The concept of an optimal portfolio can be viewed from different perspectives. From a system's point of view, flexible loads should be allocated among portfolios to maximize the total reserves, given a set of loads. However, an aggregator seeks flexible loads that increase its reserves the most. Finally, a flexible load's owner is interested in joining the portfolio in which its own contribution increases the amount of reserves of this portfolio the most. The load's view differs from the aggregator's perspective, as all aggregators want to attract reliable loads with high flexibility, but the load wishes to join only one portfolio. Here, we focus on an individual portfolio and examine how adding a flexible load affects its value. 

We consider an initial portfolio, described by the stochastic variable $X$, to which we add a flexible load. When considered in isolation, the amount of reserves that the new load can provide by itself corresponds to $q_{ \varepsilon \scriptscriptstyle Y} \left(1-R\right)$, where $\varepsilon Y$ is the stochastic variable describing the upward reserves of this load. In particular, $\varepsilon$ indicates the small amount of reserves this new load brings in comparison to the existing portfolio. The new portfolio's value equals $q_{\scriptscriptstyle X+ \scriptstyle \varepsilon \scriptscriptstyle Y} \left( 1-R \right)$. This problem can be interpreted as the change in quantile values induced by the addition of a relatively small load, which naturally aligns with the concept of a Taylor expansion. However, the small input perturbation lies in the space of probability distribution. 


The Taylor expansion in the space of distributions is called the Taylor-von Mises expansion. \added{Its exact definition is introduced in Appendix~\ref{app:vonMisesDef}.} Provided a few assumptions, e.g., the existence and derivability of a density, this expansion evaluates the evolution of statistical metrics, such as the quantile value, around an initial distribution. Based on the Taylor-Von Mises expansion as well as a few additional steps described in Appendix~\ref{app:vonMisesQaunt}, we can derive the evolution of the quantile function, in the case of a small disturbance $\varepsilon Y$ in the space of distributions, as: 
\begin{equation}
\begin{aligned}
    q_{\scriptscriptstyle X + \scriptstyle \varepsilon \scriptscriptstyle Y} \left( 1- R \right) & = q_{\scriptscriptstyle X} \left( 1-R \right) + \varepsilon \mathbb{E} \left( Y \right) \\
    & \hspace{-1.5cm} - \frac{\varepsilon^2}{2} \underbrace{\left( \ln \left( f_{\scriptscriptstyle X} \right) \right)' \left(  q_{\scriptscriptstyle X} \left(1-R \right)\right)}_{\alpha_{\scriptscriptstyle X}} \text{Var} \left( Y \right) + o\left(\varepsilon^2\right).
\end{aligned}
\label{eq:quantile_expansion}
\end{equation}
In the first order, the amount of reserves of the new portfolio increases by the expected flexibility of the new load. In the second order, it depends on the variance and derivative of the log-density of the initial portfolio at its quantile. In the following, we denote this term as $\alpha_{\scriptscriptstyle X}$. \added{In practice, computing $\alpha$ is challenging, especially based on empirical distributions. This aspect will be discussed in more detail in the next section.}

According to  (\ref{eq:quantile_expansion}), a portfolio owner should seek new flexible loads characterized by a large expected flexibility and a low variance in cases where $\alpha_{\scriptscriptstyle X}$ is positive, e.g., if the initial distribution is Gaussian. From a flexible load's point of view, its first-order contribution is identical across portfolios, but its second-order contribution is larger in the portfolio with the smallest $\alpha$ factor. Interestingly, this criterion relies on the derivative of the initial log-density at the quantile, and is, a priori, independent of the initial portfolio's size.

\section{Case Study}
\label{sec:case_study}

The considered uncertain flexible loads include EVs and HVACs.
To represent EVs, we use the ACN dataset \cite{lee_acndata_2019}, which comprises charging events recorded at three office parking-lot charging stations in California. The dataset indicates the arrival and departure times, the active charging duration, and the total energy delivered for each charging event. For some charging events, it further specifies vehicle-specific power and energy ratings, as well as the initial state-of-charge; for the others, we infer these parameters based on standard EV power and energy ratings and a plausible, randomly drawn state-of-charge. In the base case, we assume that EVs are charged at a constant power to reach their final state of charge at a user-defined departure time. While we know when EVs arrived, their exact departure times are uncertain and can be modeled as a function of arrival times using a Gaussian Mixture Model (GMM) \cite{lee_acndata_2019}. Finally, at a specific time, we only consider EVs already connected for the determination of the flexibility at future time steps and ignore those that could connect later. 

We further model the flexible electric heating systems using the large-scale dataset introduced in \cite{EGGIMANN2022111844}. This dataset provides the thermal parameters and ambient weather conditions of 300~single-family houses, for which the HVAC baseline consumption keeps the indoor temperature to 20\textdegree C. Uncertain future ambient weather conditions are represented by an auto-regressive Gaussian process, following the methodology of \cite{rousseau2026}. 


\section{Results}
\label{sec:results}

This section first illustrates the reserves that a portfolio of flexible loads can offer. Then, based on real load data, we assess the validity of our analytical reformulation of (\ref{opt:jjcc}), described in Section~\ref{sub-sub-sec:reliability_req}, and of the theoretical quantile expansion (\ref{eq:quantile_expansion}), introduced in Section~\ref{subsec:optimal_portfolio}. Finally, we investigate the existence and properties of an optimal portfolio.


\subsection{Participation of a Portfolio in the Danish Reserve Markets}

\begin{figure}
    \vspace{-0.3cm}
    \hspace{-0.4cm} 
    \includegraphics[width=\linewidth]{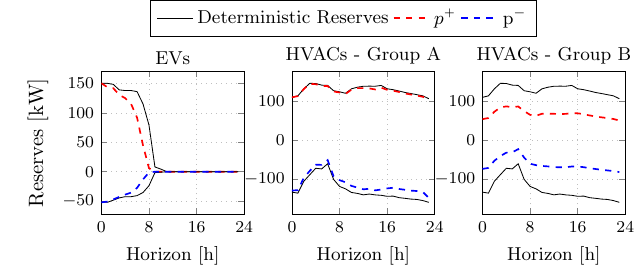}
    \vspace{-0.2cm}
    \caption{Reserves offered by different portfolios of loads over one day: a portfolio of EVs (left), a portfolio of HVACs in buildings with a wide indoor temperature comfort range (A, middle), and a portfolio of HVACs in buildings with a tight indoor temperature comfort range (B, right). Downward reserves are shown as negative for clarity in the figure.}
    \label{fig:example_B_EV}
    \vspace{-0.5cm}
\end{figure}

In Denmark, the new reserve market rules introduce energy and reliability requirements designed to open the market to uncertain, flexible loads. Fig.~\ref{fig:example_B_EV} illustrates the reserves that these loads can provide, following these new requirements. More specifically, it depicts the reserve capability of multiple EVs and HVACs over a winter day, starting at 08:00. Deterministic reserves, i.e., in the absence of uncertainty, are shown by solid lines, while reserves in the presence of uncertainty and computed using (\ref{eq:quantile_decoupled}) are shown by dashed lines. Over the first 4~hours, it is likely that EVs remain connected, as they are connected at a workplace. Hence, the dashed and solid lines superpose. Afterwards, the probability that EVs disconnect increases, hence, the level of reserves they can offer with a reliability of 90\% decreases. Eventually, EVs are disconnected, and the reserves fall to zero. In contrast, portfolios composed of HVACs provide a stable level of reserves throughout the day. Fig.~\ref{fig:example_B_EV} depicts the reserves of two groups of HVACs. In Group A, HVACs serve buildings with a large acceptable range of indoor temperatures, defined as $\left[ 18.5, 21.5\right]$\textdegree C, while in Group B, these buildings are characterized by a tighter range of $\left[ 19, 21\right]$\textdegree C. Additionally, we assume that Group B is exposed to more uncertain future weather conditions than Group A. As a consequence, in Group B, the uncertain weather conditions have a greater impact, reducing the available flexibility. 

\begin{figure}
    \hspace{-0.4cm} \includegraphics[width=\linewidth]{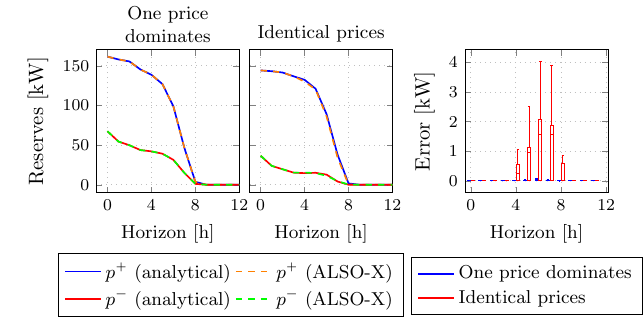}
    \caption{Comparison between the proposed analytical method and the ALSO-X method, for an example day, when one reserve price dominates (left) and when both are equal (middle), and the distribution of absolute errors over 20~days (right). }
    \label{fig:comparison_alsox}
    \vspace{-0.3cm}
\end{figure}


\subsection{Validation of the Analytical Reformulation}

In Section~\ref{sub-sub-sec:reliability_req}, we reformulated the joint chance-constrained optimization (\ref{opt:jjcc}) into the analytical problem (\ref{eq:quantile_decoupled}). To validate our approach, we benchmark our analytical method against the ALSO-X method\footnote{A brief description of the ALSO-X formulation is provided in Appendix~\ref{app:also_x}.}, an established approach for solving joint chance-constrained problems.
Fig.~\ref{fig:comparison_alsox} compares the amount of reserves that a portfolio of EVs can offer, according to both methods. When one reserve price dominates, Section~\ref{sub-sub-sec:reliability_req} introduces an exact analytical reformulation of the problem, as Fig.~\ref{fig:comparison_alsox} illustrates. At the worst hour, we observe a mean absolute error of 0.0195~kW\added{, a 75\%-quantile of 0.035~kW, and a maximum absolute error of 0.063~kW} between the two methods. \added{This small difference stems from the sample-based quantile estimation, which introduces small numerical errors. Additionally, the computation time for reserve scheduling over a day drops from around 9~minutes with the ALSO-X method to around 1.2~seconds with our analytical reformulation, on average.} When both prices are equal, the analytical reformulation is not exact, as Fig.~\ref{fig:comparison_alsox} shows. Yet, the mean absolute error between the two methods remains small relative to the reserve magnitude, with a mean error of 1.32~kW\added{, a 75\%-quantile of 2.06~kW, and a maximum absolute error of 4.03~kW} at the worst hour, which should be acceptable in practice.


\subsection{Validation of the Quantile Expansion}
\label{subsec:quantile_exp_validation}

To study the notion of an optimal portfolio, we derived the quantile expansion (\ref{eq:quantile_expansion}), approximating the marginal change in reserve capacity when a new flexible load is added to an existing portfolio. To empirically validate this formula, we evaluate the increase in reserves resulting from adding various new loads to an existing EV portfolio. More specifically, we generate 200~loads whose flexibilities are governed by a randomly assigned probability distribution, selected from a set of standard probability distributions, with randomly selected parameters. According to (\ref{eq:quantile_expansion}), the marginal change in reserves can be approximated by the new load's expected flexibility, in the first order term, which is confirmed in the results shown in Fig.~\ref{fig:first_and_second_order}. In the second order term, it linearly depends on the variance of the new load, which is also validated in Fig.~\ref{fig:first_and_second_order}. 

\begin{figure}
    \vspace{-0.2cm}
    \hspace{0.2cm}
    \includegraphics[width=0.9\linewidth]{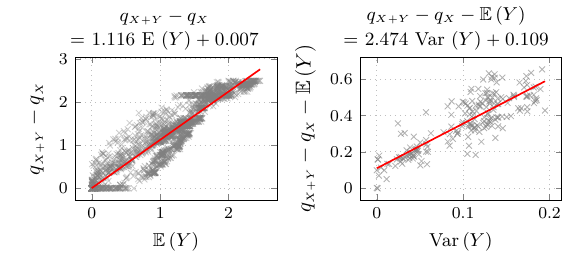}
    \vspace{-0.2cm}
    \caption{Validation of (\ref{eq:quantile_expansion}) for the first order (left) and the second order terms (right).}
    \vspace{-0.6cm}
    \label{fig:first_and_second_order}
\end{figure}


The second-order term depends not only on the new load's variance but also on the derivative of the log-density of the initial portfolio evaluated at its quantile. However, this quantity is very challenging to compute using sampled data. It also varies across time and must be estimated for every hour. To overcome this issue, we leverage the correlation observed between the marginal change in reserves and the new load's variance in the second order term. For each hour, we evaluate the correlation and use it to infer the log-density derivative. This approach is used in the rest of the paper.

\vspace{-0.1cm}

\subsection{Optimal Portfolio of Flexible Loads}
\begin{figure}
    \vspace{-0.2cm}
    \centering
    \includegraphics[width=\linewidth]{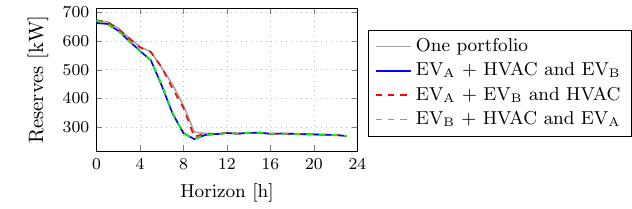}
    \vspace{-0.6cm}
    \caption{Total amount of reserves obtained with two groups of EVs, and one group of 200~buildings, for different pairings into portfolios.}
    \label{fig:example_combination2}
    \vspace{-0.3cm}
\end{figure}


The existence of an optimal portfolio suggests that appropriately combining flexible loads can increase the total reserve capacity. To illustrate this effect, we consider an example in which three groups of loads are combined into two portfolios. Specifically, we analyze two groups of EVs, each composed of approximately 50~vehicles and one group of HVACs, comprising 200~HVACs with a wide temperature comfort range, similar to Group A depicted in Fig.~\ref{fig:example_B_EV}. For the three possible configurations, Fig.~\ref{fig:example_combination2} depicts the total reserves that both portfolios can offer, as well as reserves that loads can offer when all grouped into one portfolio. One configuration clearly dominates the others, namely when the two EV groups are paired together. The HVAC group is highly reliable, as shown in Fig.~\ref{fig:example_B_EV}. Hence, when paired with an EV group, its reliability balances the uncertainty of one of the EV groups, but the other group of EVs is not reliable enough to provide substantial reserves. In contrast, pairing the two EV groups improves their total reliability while allowing the HVAC group to independently deliver a high reserve capacity. \added{At the worst hour, pairing each EV group with the HVAC group reduces total reserve capacity by 25.6\% relative to the single-portfolio benchmark, while pairing the two EV groups together narrows this gap to 5.8\%.} This example indicates that optimal portfolio formation can be interpreted as an optimal allocation of a reliability budget.


\begin{figure}
    \hspace{-0.4cm}
    \includegraphics[width=1.1\linewidth]{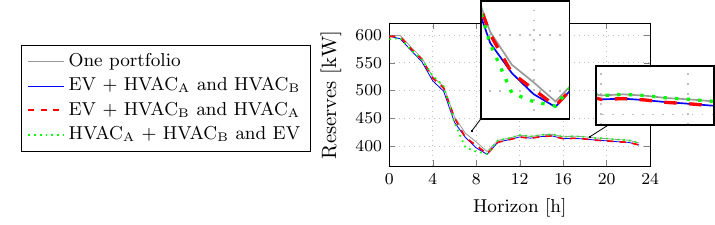}
    \vspace{-0.6cm}
    \caption{Total amount of reserves obtained with a group of EVs, and two different groups of 200~buildings each, for different pairings into portfolios.}
    \label{fig:example_combination}
    \vspace{-0.6cm}
\end{figure}

Figure~\ref{fig:example_combination} presents the results of a similar experiment with one EV group and two HVAC groups A and B, introduced in Fig.~\ref{fig:example_B_EV}. In contrast to the previous example, the optimal configuration varies over time. Between hours 4 and 8, when the EV uncertainty is high, pairing EVs with an HVAC group is beneficial, with a slight preference for the more uncertain HVAC group. \added{At the worst hour, the optimal single-portfolio configuration offers 6\% more reserves than the worst alternative pairing, with the remaining combinations falling 1.8\% and 2.6\% short.} After hour~12, EVs are no longer available, and pairing them yields no additional reliability or value. In this case, the two HVAC groups are optimally paired. \added{This example highlights that the optimal portfolio composition is not static, but evolves as the uncertainty profiles of the available loads change throughout the day.}


\begin{figure}
    \vspace{-0.2cm}
    \centering
    \includegraphics[width=0.9\linewidth]{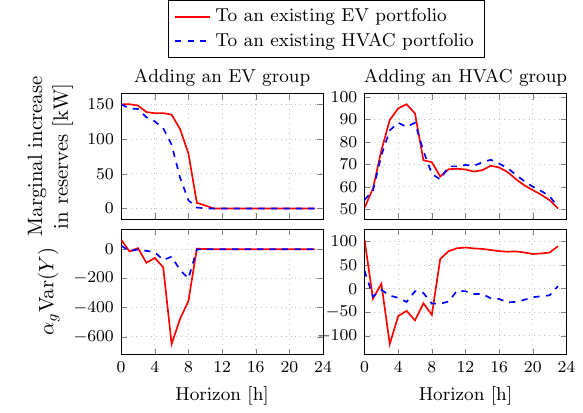}
    \vspace{-0.2cm}
    \caption{Marginal increase in reserves resulting from adding different loads to different portfolios.}
    \label{fig:added_value_analyis}
    \vspace{-0.5cm}
\end{figure}


Flexible loads seeking to join an existing portfolio can leverage the proposed methodology to identify the portfolio that would benefit most from their participation. Indeed, Fig.~\ref{fig:added_value_analyis} shows the marginal increase in reserves resulting from adding a group of EVs or a group of HVACs to different portfolios. For the EV group, joining an EV-only portfolio yields a higher marginal increase in reserves, whereas the HVAC group may prefer joining different portfolios at different times of the day. According to the quantile expansion (\ref{eq:quantile_expansion}), flexible loads should choose to join portfolios for which the product of the log-density derivative, $\alpha$, and their variance is the smallest. Using the data-based identification of the log-density derivative described earlier, Fig.~\ref{fig:added_value_analyis} shows that, by following this criterion, flexible loads can identify which portfolio to join.

\added{Figures~\ref{fig:example_combination} and~\ref{fig:example_combination2} suggest that grouping all loads into a single portfolio tends to yield better results than splitting them across multiple portfolios. We do not prove this holds in general, and counterexamples may exist, for instance when loads are strongly correlated. In practice, multiple portfolios often emerge naturally due to contractual or operational constraints, rather than by design. Our results nonetheless quantify the optimality gap between such fragmented configurations and the single-portfolio benchmark, which may serve as a useful bound for practitioners.}






\section{Conclusion}
\label{sec:conclusion}


This paper proposes an analytical framework to quantify the reserve capacity of portfolios composed of energy-constrained and uncertain flexible loads. The method uses a quantile-based formulation of a portfolio's flexibility, which is exact in some specific cases and results in small approximation errors in most practical settings. Yet, this method is not only a computationally efficient way to compute reliable reserves, but it can also be used to analyze the factors that affect a portfolio's reserves. In particular, it allows us to understand how the marginal contribution of an additional load varies with the stochastic properties of the loads. Based on a few selected examples, we demonstrate the existence of an optimal load grouping, given a fixed number of loads and portfolios. We additionally show that our framework helps to identify portfolios in which individual loads deliver the highest marginal value.

Future work should extend the initial results presented in this paper to larger-scale case studies and larger portfolio sizes. Furthermore, the current theoretical framework assumes that loads are affected by independent sources of uncertainty, which may not hold in practice. Accounting for correlated uncertainties is necessary to extend our methodology to more realistic case studies. 


\section{AI Usage Disclosure}

We acknowledge the use of generative AI for spell- and grammar-checking, and style improvements.

\appendix 
\subsection{\added{Optimal Portfolio: An Illustrative Example}} 
\label{app:example}

\added{To illustrate the concept of optimal portfolio formations, we introduce a simple example. As indicated in Table~\ref{tab:illustrativeExample}, we consider 4~loads, labeled A to D. We assume that each load can provide a discretized amount of upward flexibility, with each discrete level associated with a reliability. For instance, Load A can increase its consumption by 0.5~kW with a confidence of 70\%, and by 1~kW with a confidence of 30\%. }

\begin{table}[t]
    \centering
    \footnotesize
    \renewcommand{\arraystretch}{1.3}
    \setlength{\tabcolsep}{3pt}
    \caption{Example of 4~flexible loads with 2~levels of flexibility (left) and possible splits into 2~portfolios of 2~loads each (middle and right).}
    \vspace{-0.3cm}
    \label{tab:illustrativeExample}
    \hspace{-0.2cm}
    \begin{minipage}{0.28\columnwidth}
    \hspace{0.7cm} $\scriptstyle \mathbb{P}\left( \mathrm{Reserve} \right) = x$ \par
    \vspace{0.3em}
    \centering 
    \begin{tabular}{|c|c|c|}
        \hline
        $x$ & 0.5~kW & 1~kW \\
        \hline
        \textbf{A} & 70\% & 30\% \\
        \textbf{B} & 70\% & 30\% \\
        \textbf{C} & 50\% & 50\% \\
        \textbf{D} & 50\% & 50\% \\
        \hline
        \end{tabular}
    \end{minipage}%
    \begin{minipage}{0.43\columnwidth}
        \hspace{1.6cm} $\scriptstyle \mathbb{P}\left( \mathrm{Reserve} \right) = x$  \par
        \vspace{0.3em}
        \centering
        \begin{tabular}{|c|c|c|c|}
        \hline
        $x$ & 2~kW & 1.5~kW & 1~kW \\
        \hline
        \textbf{A \& B} & 9\% & 42\% & 49\% \\
        \textbf{C \& D} & 25\% & 50\% & 25\% \\
        \hline
        \textbf{A \& C} & 15\% & 50\% & 35\% \\
        \textbf{B \& D} & 15\% & 50\% & 35\% \\
        \hline
    \end{tabular}
    \end{minipage}
    \begin{minipage}{0.28\columnwidth}
        \centering
        $\scriptstyle \mathbb{P} \left( \mathrm{Reserve} \leq x \right) \geq 70\%$\par
        \vspace{0.3em}
        \begin{tabular}{|c|c|}
        \hline
        \multirow{2}{*}{} & Reserves \\
        \multirow{2}{*}{} & at 70\% ($x$) \\
        \hline
        \textbf{A \& B} & 1~kW\\
        \textbf{C \& D} & 1.5~kW\\
        \hline
        \textbf{A \& C} & 1~kW\\
        \textbf{B \& D} & 1~kW\\
        \hline
    \end{tabular}
    \end{minipage}
    \vspace{-0.5cm}
\end{table}

\added{Assuming that we want to form two portfolios, each containing two loads, there exist two possible groupings: we form either homogeneous or heterogeneous groups. Table~\ref{tab:illustrativeExample} indicates the reliability associated with different total upward reserve values for each group. For instance, if Loads A and B are paired, their total consumption can be increased by 2~kW if each load increases its individual consumption by 1~kW, with a reliability of $30\%\times 30\% = 9\%$. Similarly, 1~kW of total reserves corresponds to both delivering 0.5~kW of upward reserves, which is associated with a probability of 49\%. Finally, Loads A and B can increase their total consumption by 1.5~kW with a reliability of 42\%.} 

\added{Loads A and B can provide at least 1.5~kW of reserves with confidence 9\%+42\% = 51\% and at least 1~kW with full certainty. Hence, given a reliability level of 70\%, they can reliably provide 1~kW of upwards reserves. Similarly, analyzing the cumulative distribution of reserves of Loads C and D, we observe that they can deliver an amount of reserves larger than 1.5~kW with a 70\% confidence. Indeed, for the second portfolio, the probability that the reserves are larger than 1.5~kW is 75\%, exceeding the 70\%-reliability requirement. The value of 1.5~kW also corresponds to the 30\%-quantile of the discrete stochastic variable describing the reserves of Loads C and D, as introduced in Section~\ref{sub-sub-sec:reliability_req}. When grouping loads heterogeneously, each group can only provide 1~kW of upward reserves, under the 70\%-requirement, as indicated in Table~\ref{tab:illustrativeExample}. Hence, under these conditions, the total reserve level is higher with homogeneous grouping, indicating that, for a fixed set of resources and number of portfolios, an optimal portfolio allocation exists. }

\subsection{ALSO-X Formulation}
\label{app:also_x}

Using a scenario approach, the joint chance-constrained problem (\ref{opt:jjcc}) can be reformulated as: 
\begin{subequations}
    \begin{align}
        \min_{p^+_h, p^-_h} & \quad c^+_h p^+_h + c^-_h p^-_h \\
        \text{s.t. } \quad & p_{i,h, s}^+ + 0.2 p_{i,h,s}^- \leq p^{\mathrm{up},+}_{i,h,s}, \quad \forall i \in \mathcal{I}, \forall s \in \mathcal{S}, \label{opt-cstr:app_1}\\
         & p_{i,h,s}^- + 0.2 p_{i,h,s}^+ \leq p^{\mathrm{up},-}_{i,h,s}, \quad \forall i \in \mathcal{I}, \forall s \in \mathcal{S}, \label{opt-cstr:app_2}\\
         & 0 \leq p^+_{i,h,s} \leq e^{\mathrm{up},+}_{i,h,s}, \hspace{1.35cm} \forall i \in \mathcal{I}, \forall s \in \mathcal{S}, \label{opt-cstr:app_3}\\
         & 0 \leq p^-_{i,h,s} \leq e^{\mathrm{up},-}_{i,h,s}, \hspace{1.35cm} \forall i \in \mathcal{I}, \forall s \in \mathcal{S}, \label{opt-cstr:app_4} \\
         & p^+_h \leq \sum_{i\in \mathcal{I}} p_{i,h,s}^+ + My_{h,s}, \hspace{1.55cm} \forall s \in \mathcal{S}, \label{opt-cstr:app_5} \\
         & p^-_h \leq \sum_{i\in \mathcal{I}} p_{i,h,s}^- + My_{h,s}, \hspace{1.55cm} \forall s \in \mathcal{S}, \label{opt-cstr:app_6} \\ 
         & y_{h,s} \in \{0, 1\}, \hspace{3.3cm} \forall s \in \mathcal{S}, \label{opt-cstr:app_7} \\
         & \frac{1}{\left| \mathcal{S} \right|} \sum_{s\in \mathcal{S}} y_{h,s} \leq 1-R, \label{opt-cstr:app_8}
    \end{align}
    \label{opt:app}
\end{subequations}
where the binary variables $y_{h,s}$ and (\ref{opt-cstr:app_8}) allow the violation of constraints in at most $(1-R)$\% of the scenarios contained in the set $\mathcal{S}$. The ALSO-X method relaxes this problem by considering the variables $y_{h,s}$ as continuous. In return, Constraint (\ref{opt-cstr:app_8}) is transformed into a budget constraint as: 
\begin{equation}
    \sum_{s\in \mathcal{S}} y_{h,s} \leq Q, 
\end{equation}
where the violation budget $Q$ is iteratively changed until only $(1-R)$\% of the scenarios are violated. A more rigorous definition of the problem can be found in \cite{Lunde2025}.

\subsection{Von Mises-Taylor Expansion of the Quantile Function}
\label{app:expansion}

\subsubsection{Statistical Functions}
Von Mises' theory relies on the observation that statistical functions, such as the quantile function, can be seen as functions of a stochastic cumulative distribution \cite{fernholz1983vonmises}. Mathematically, this means that: 
\begin{equation}
    q_{\scriptscriptstyle X} \left( 1-R \right) = T \left( F_{\scriptscriptstyle X} \right),
\end{equation}
where $F_{\scriptscriptstyle X}$ is the cumulative density function of a stochastic variable $X$ and $T$ is defined for a fixed reliability $R$.

\subsubsection{Von Mises-Taylor Expansion}
\label{app:vonMisesDef}
Based on the concept of statistical functions, Von Mises \textit{"developed a theory for the analysis of the asymptotic distribution of statistical [functions], using a form of Taylor expansion involving the derivatives of the [statistical functions]"} \cite{fernholz1983vonmises}. In other words, Von Mises developed an equivalent to the Taylor expansion, but for statistical functions in the space of probability distributions. 

A prerequisite to the Von Mises expansion is the definition of the derivative of a statistical function in the space of probability distributions. The Von Mises derivative of a statistical function $T$ is defined if there exists a function $\phi$ such that: 
\begin{equation}
    \left. \frac{d}{dt} T \left( F_{\scriptscriptstyle X} + t \left( F_{\scriptscriptstyle Z} - F_{\scriptscriptstyle X} \right) \right) \right|_{t = 0} = \int \phi \left( F_{\scriptscriptstyle X}, u \right) \text{d} \left( F_{\scriptscriptstyle Z} - F_{\scriptscriptstyle X} \right) (u).
\end{equation}
From this definition, we can see that the derivative is directional, i.e., it depends on the chosen direction $F_{\scriptscriptstyle Z}$. In the following, we will denote the derivative of $T$, at $F_{\scriptscriptstyle X}$, in direction $F_{\scriptscriptstyle Z}$, as $T'\left( F_{\scriptscriptstyle X}; F_{\scriptscriptstyle Z} \right)$. Additionally, we will use the following notation: 
\begin{equation}
    F_{{\scriptscriptstyle X}, t}  = F_{\scriptscriptstyle X} + t \left( F_{\scriptscriptstyle Z} - F_{\scriptscriptstyle X} \right) = (1-t) F_{\scriptscriptstyle X} + t F_{\scriptscriptstyle Z}.
\end{equation}

Under the assumption that $T$ is twice-differentiable in $F_{\scriptscriptstyle X}$, the Von Mises-Taylor expansion at the second order is:
\begin{equation}
\begin{aligned}
    T \left(F_{\scriptscriptstyle Z} \right) & - T \left( F_{\scriptscriptstyle X} \right) = T' \left( F_{\scriptscriptstyle X}; F_{\scriptscriptstyle Z} \right) \\ 
    & + \frac{1}{2} T'' \left( F_{\scriptscriptstyle X}; F_{\scriptscriptstyle Z} \right) + \text{Rem} \left(F_{\scriptscriptstyle X} - F_{\scriptscriptstyle Z} \right),
\end{aligned}
    \label{eq:app_vonMisesExpansion}
\end{equation}
where $\text{Rem} \left(F_{\scriptscriptstyle X} - F_{\scriptscriptstyle Z} \right)$ designates a remainder term. 

\subsubsection{Application to the Quantile Function}
\label{app:vonMisesQaunt}
Developing the Von Mises-Taylor expansion of the quantile function requires first demonstrating that the quantile function, as a statistical function, is twice differentiable. Starting with the first-order derivative, the authors of \cite{bartlett2013theoretical} prove its differentiability and find the first derivative's value using the following property of the quantile function: 
\begin{equation}
    F_{{\scriptscriptstyle X},t} \left( T\left(F_{{\scriptscriptstyle X}, t} \right) \right) = 1-R,
\end{equation}
such that: 
\begin{equation}
    \frac{d}{dt} F_{{\scriptscriptstyle X},t} \left( T\left(F_{{\scriptscriptstyle X},t} \right) \right) = 0.
\end{equation}
Based on this equality, we apply the chain rule and obtain:
\begin{equation}
\begin{aligned}
    & \left. \frac{d}{dt} \left( (1-t)F_{\scriptscriptstyle X} \left(T \left( F_{{\scriptscriptstyle X},t} \right) \right) + t F_{\scriptscriptstyle Z} \left( T \left( F_{{\scriptscriptstyle Z},t} \right) \right) \right) \right|_{t=0} \\
    & =  \left[ -F_{\scriptscriptstyle X} \left( T \left(F_{{\scriptscriptstyle X},t} \right)\right) + (1-t) F_{\scriptscriptstyle X}'\left(T \left( F_{{\scriptscriptstyle X}, t} \right)\right) \frac{d}{dt} T(F_{{\scriptscriptstyle X},t})  \right.\\
    & \hspace{1cm} \left. \left. + F_{\scriptscriptstyle Z} \left( T \left( F_{{\scriptscriptstyle X},t} \right) \right) + t F'_{\scriptscriptstyle Z} \left(T \left(F_{{\scriptscriptstyle X},t} \right) \right)\frac{d}{dt} T\left(F_{{\scriptscriptstyle X},t}\right) \right] \right|_{t=0} \\
    & = -F_{\scriptscriptstyle X} \left(T \left(F_{\scriptscriptstyle X}\right) \right) + F_{\scriptscriptstyle X}' \left( T\left(F_{\scriptscriptstyle X}\right) \right) T'\left( F_{\scriptscriptstyle X}; F_{\scriptscriptstyle Z} \right) + F_{\scriptscriptstyle Z} \left( T\left(F_{\scriptscriptstyle X}\right) \right) \\
    & = 0.
\end{aligned}
\label{eq:derivation_firstorder_derivative}
\end{equation}
Eventually, we can conclude that:
\begin{equation}
    T'\left( F_{\scriptscriptstyle X}; F_{\scriptscriptstyle Z} \right) = \frac{F_{\scriptscriptstyle X} \left(T \left(F_{\scriptscriptstyle X} \right) \right) - F_{\scriptscriptstyle Z} \left( T \left(F_{\scriptscriptstyle X}\right) \right)}{F_{\scriptscriptstyle X}'\left( T \left(F_{\scriptscriptstyle X} \right) \right)}.
\end{equation}
Similarly, we can derive the second-order derivative. Using the same approach as in (\ref{eq:derivation_firstorder_derivative}), we compute: 
\begin{equation}
\begin{aligned}
    & \left. \frac{d^2}{dt^2} \left( (1-t)F_{\scriptscriptstyle X} \left( T \left( F_{{\scriptscriptstyle X},t} \right) \right) + t F_{\scriptscriptstyle Z} \left( T \left( F_{{\scriptscriptstyle X}, t} \right) \right) \right) \right|_{t=0} \\
    & = \frac{d}{dt} \left[ -F_{\scriptscriptstyle X} \left( T \left( F_{{\scriptscriptstyle X}, t} \right)\right) \right. \\
    & \hspace{0.5cm} + (1-t) F_{\scriptscriptstyle X}'\left( T \left(F_{{\scriptscriptstyle X},t} \right)\right) \frac{d}{dt} T \left(F_{{\scriptscriptstyle X},t} \right) \\
    & \hspace{0.5cm} \left. \left. + F_{\scriptscriptstyle Z} \left( T \left( F_{{\scriptscriptstyle X}, t} \right) \right) + t F_{\scriptscriptstyle Z}'\left( T \left( F_{{\scriptscriptstyle X}, t} \right) \right)\frac{d}{dt} T \left( F_{{\scriptscriptstyle X},t} \right) \right] \right|_{t=0} \\
    & = \left[ -F_{\scriptscriptstyle X}'\left(T \left(F_{{\scriptscriptstyle X},t} \right)\right) \frac{d}{dt} T \left( F_{{\scriptscriptstyle X},t} \right) \right. \\
    & \hspace{0.5cm} - F_{\scriptscriptstyle X}'\left( T \left( F_{{\scriptscriptstyle X}, t} \right) \right) \frac{d}{dt} T \left( F_{{\scriptscriptstyle X},t} \right)  \\
    & \hspace{0.5cm} + (1-t) F_{\scriptscriptstyle X}''\left( T \left( F_{{\scriptscriptstyle X},t} \right)\right) \left(\frac{d}{dt} T \left( F_{{\scriptscriptstyle X},t} \right) \right)^2 \\
    & \hspace{0.5cm} + (1-t) F_{\scriptscriptstyle X}' \left(T \left( F_{{\scriptscriptstyle X},t} \right)\right) \frac{d^2}{dt^2} T \left( F_{{\scriptscriptstyle X},t} \right) \\
    & \hspace{0.5cm} + F_{\scriptscriptstyle Z}'\left( T \left( F_{{\scriptscriptstyle X},t} \right) \right) \frac{d}{dt} T\left( F_{{\scriptscriptstyle X}, t} \right) \\
    & \hspace{0.5cm} + F_{\scriptscriptstyle Z}'\left( T \left( F_{{\scriptscriptstyle X},t} \right) \right)\frac{d}{dt} T \left( F_{{\scriptscriptstyle X},t} \right) \\
    & \hspace{0.5cm}+ t F_{\scriptscriptstyle Z}''\left( T \left(F_{{\scriptscriptstyle X},t} \right) \right) \left(\frac{d}{dt} T \left(F_{{\scriptscriptstyle X}, t} \right) \right)^2 \\
    & \left. \left. \hspace{0.5cm} + t F_{\scriptscriptstyle Z}' \left( T \left(F_{{\scriptscriptstyle X},t} \right) \right)\frac{d^2}{dt^2} T \left( F_{{\scriptscriptstyle X}, t} \right) \right] \right|_{t=0} \\
    & = 2 \left[ F_{\scriptscriptstyle Z}'\left( T \left(F_{\scriptscriptstyle X} \right)\right) - F_{\scriptscriptstyle X}'\left( T \left( F_{\scriptscriptstyle X} \right)\right) \right] T' \left(F_{\scriptscriptstyle X}; F_{\scriptscriptstyle Z} \right) \\
    & \hspace{0.5cm}  + F_{\scriptscriptstyle X}'' \left( T \left(F_{\scriptscriptstyle X} \right)\right) \left( T' \left(F_{\scriptscriptstyle X}; F_{\scriptscriptstyle Z} \right) \right)^2 \\
    & \hspace{0.5cm} + F_{\scriptscriptstyle X}'\left( T \left(F_{\scriptscriptstyle X} \right)\right) T'' \left(F_{\scriptscriptstyle X}; F_{\scriptscriptstyle Z} \right) \\
    &= 0.
\end{aligned}
\label{eq:derivation_secondorder_derivative}
\end{equation}
Finally, we obtain: 
\begin{equation}
\begin{aligned}
    T'' \left(F_{\scriptscriptstyle X}; F_{\scriptscriptstyle Z} \right) & = \frac{2 \left( F_{\scriptscriptstyle X}'\left( T \left( F_{\scriptscriptstyle X} \right)\right) - F_{\scriptscriptstyle Z}'\left( T \left( F_{\scriptscriptstyle X} \right)\right) \right) T' \left(F_{\scriptscriptstyle X}; F_{\scriptscriptstyle Z}\right)}{F_{\scriptscriptstyle X}'\left(T \left(F_{\scriptscriptstyle X} \right)\right)} \\ 
    & \hspace{0.5cm} - \frac{F''_{\scriptscriptstyle X} \left(T \left( F_{\scriptscriptstyle X} \right) \right) \left(T' \left( F_{\scriptscriptstyle X}; F_{\scriptscriptstyle Z} \right) \right)^2}{F'_{\scriptscriptstyle X}\left( T \left( F_{\scriptscriptstyle X} \right)\right)}.
\end{aligned}
\end{equation}
Replacing the first and second-order derivative formulae in (\ref{eq:app_vonMisesExpansion}), we obtain:  
\begin{equation}
    \begin{aligned}
        q_{\scriptscriptstyle Z}  - q_{\scriptscriptstyle X}  & = \frac{F_{\scriptscriptstyle X} \left(q_{\scriptscriptstyle X} \right) - F_{\scriptscriptstyle Z} \left(q_{\scriptscriptstyle X} \right)}{f_{\scriptscriptstyle X} \left(q_{\scriptscriptstyle X} \right)} \\
        & + \frac{\left( f_{\scriptscriptstyle X} \left( q_{\scriptscriptstyle X} \right) - f_{\scriptscriptstyle Z} \left(q_{\scriptscriptstyle X} \right) \right) \left(F_{\scriptscriptstyle X} \left( q_{\scriptscriptstyle X} \right) - F_{\scriptscriptstyle Z} \left( q_{\scriptscriptstyle X} \right) \right)}{f_{\scriptscriptstyle X} \left( q_{\scriptscriptstyle X} \right)^2} \\
        & - \frac{f_{\scriptscriptstyle X}' \left(q_{\scriptscriptstyle X}\right) \left( F_{\scriptscriptstyle X} \left( q_{\scriptscriptstyle X} \right) - F_{\scriptscriptstyle Z} \left( q_{\scriptscriptstyle X} \right) \right)^2}{2 f_{\scriptscriptstyle X} \left( q_{\scriptscriptstyle X} \right)^3},
    \end{aligned}
    \label{eq:app_before_exp}
\end{equation}
where the simplified notation $q_{\scriptscriptstyle X}$ designates $q_{\scriptscriptstyle X} \left( 1-R \right)$.

In the optimal portfolio problem, we wish to approximate the quantile of the portfolio $Z = X + \varepsilon Y$, where $X$ represents the initial portfolio and $\varepsilon Y$ the added load. To this aim, according to (\ref{eq:app_before_exp}), we should evaluate the cumulative density function $Z$, in $q_{\scriptscriptstyle X}$. However, the existing portfolio would not reasonably share its stochastic profile with the new load and vice versa when prospecting for new loads. In practice, we could evaluate some stochastic properties of $X$ or $\varepsilon Y$, but not $Z$. To solve this issue, one approach is to linearize the cumulative distribution function $F_{\scriptscriptstyle Z}$ around $F_{\scriptscriptstyle X}$ using a Taylor expansion. If $X$ and $\varepsilon Y$ are independent, we can prove that:
\begin{equation}
    F_{\scriptscriptstyle Z} (u) = F_{\scriptscriptstyle X+ \scriptstyle \varepsilon \scriptscriptstyle Y} (u) = \mathbb{E} \left[ F_{\scriptscriptstyle X} \left(u - \varepsilon Y \right) \right].
\end{equation}
Therefore, we first linearize $F_{\scriptscriptstyle X}$ around $u$ and, then, consider the expected value of the expression to obtain: 
\begin{equation}
\begin{aligned}
    F_{\scriptscriptstyle X} & \left( u \right) - F_{\scriptscriptstyle Z} (u) = \varepsilon \mathbb{E} \left( Y \right) f_{\scriptscriptstyle X} \left( u \right) \\
    & - \frac{\varepsilon^2}{2} \mathbb{E} \left( Y^2 \right) f_{\scriptscriptstyle X} '(u) + o \left( \varepsilon^2 \right).
\end{aligned}
\end{equation}
Considering $u = q_{\scriptscriptstyle X}$, we can simplify (\ref{eq:app_before_exp}) to obtain the following second-order expression:
\begin{equation}
    \begin{aligned}
        q_{\scriptscriptstyle Z} & - q_{\scriptscriptstyle X} = \varepsilon \mathbb{E} \left( Y \right) - \frac{\varepsilon^2}{2} \mathbb{E} \left(Y^2 \right) \frac{f_{\scriptscriptstyle X}' (q_{\scriptscriptstyle X})}{f_{\scriptscriptstyle X} (q_{\scriptscriptstyle X} )} \\
        & + \varepsilon^2 \mathbb{E} (Y)^2  \frac{f_{\scriptscriptstyle X}' (q_{\scriptscriptstyle X})}{f_{\scriptscriptstyle X} (q_{\scriptscriptstyle X} )} - \frac{\varepsilon^2}{2} \mathbb{E} (Y)^2 \frac{f_{\scriptscriptstyle X}' (q_{\scriptscriptstyle X})}{f_{\scriptscriptstyle X} (q_{\scriptscriptstyle X} )} + o \left( \varepsilon^2 \right) \\
        & \hspace{-0cm} = \varepsilon \mathbb{E} \left( Y \right) - \frac{\varepsilon^2}{2} \frac{f_{\scriptscriptstyle X}' (q_{\scriptscriptstyle X})}{f_{\scriptscriptstyle X} (q_{\scriptscriptstyle X} )} \left( \mathbb{E} \left(Y^2 \right) - \mathbb{E} (Y)^2 \right) + o \left( \varepsilon^2 \right)  \\
        & \hspace{-0cm} = \varepsilon \mathbb{E} \left( Y \right) - \frac{\varepsilon^2}{2} \frac{f_{\scriptscriptstyle X}' (q_{\scriptscriptstyle X})}{f_{\scriptscriptstyle X} (q_{\scriptscriptstyle X} )} \text{Var} (Y) + o \left( \varepsilon^2 \right).
    \end{aligned}
    \label{eq:finalExpansionDerivation}
\end{equation}

Finally, in (\ref{eq:app_vonMisesExpansion}), we only consider a second-order expansion. As a consequence, there is a remainder term which could introduce first and second-order $\varepsilon$ terms. Hence, to complete this development, we should prove that the remainder term in (\ref{eq:app_vonMisesExpansion}) is $o(\varepsilon^2)$. To this aim, we should demonstrate that the $k^\mathrm{th}$ derivative of $T$ evaluated at $F_{\scriptscriptstyle X}$ in direction $F_{\scriptscriptstyle Z}$ exists and is of order $\varepsilon^k$. This can be proven by recurrence, using a similar approach to that for the first- and second-order derivatives. However, for brevity, we omit the detailed proof here.

\subsubsection{Verification in the Gaussian Case}

 We want to verify (\ref{eq:quantile_expansion}) in the Gaussian case, for which there exists an explicit formulation of the quantile function. If a stochastic variable $X$ is Gaussian, we can relate its $(1-R)$-quantile to the $(1-R)$-quantile of the centered standard Gaussian, $q_{0,1}$, as: 
 \begin{equation}
     q_{\scriptscriptstyle X} = \mu_{\scriptscriptstyle X} + \sqrt{ \sigma_{\scriptscriptstyle X}^2} q_{0,1}, 
     \label{eq:app_gaussianX_quantile}
 \end{equation}
 where $\mu_{\scriptscriptstyle X}$ and $\sigma_{\scriptscriptstyle X}^2$ denote the mean and variance of $X$, respectively. Similarly, considering $Z = X+\varepsilon Y$, where $X$ and $Y$ are two independent Gaussian variables, we know that $Z$ is Gaussian and its $(1-R)$-quantile verifies: 
 \begin{equation}
     q_{\scriptscriptstyle Z} = \mu_{\scriptscriptstyle X} + \varepsilon \mu_{\scriptscriptstyle Y} + \sqrt{\sigma_{\scriptscriptstyle X}^2 + \varepsilon^2 \sigma_{\scriptscriptstyle Y}^2} q_{0,1}.
     \label{eq:app_gaussian_XY_quantile}
 \end{equation}
The change of $(1-R)$-quantile resulting from adding $\varepsilon Y$ to $X$ is: 
 \begin{equation}
     q_{\scriptscriptstyle Z}  - q_{\scriptscriptstyle X}  = \varepsilon \mu_{\scriptscriptstyle Y} + \underbrace{\left ( \sqrt{\sigma_{\scriptscriptstyle X}^2 + \varepsilon^2 \sigma_{\scriptscriptstyle Y}^2} - \sqrt{\sigma_{\scriptscriptstyle Z}^2} \right)}_{\Delta_\sigma} q_{0,1} . 
     \label{eq:app_gaussian_case_comparison}
 \end{equation}
Considering that $\varepsilon$ is small, we can linearize $\Delta_\sigma$ around $\sigma_{\scriptscriptstyle X}^2$, such that, at the second order, we obtain:
\begin{equation}
    q_{\scriptscriptstyle Z} - q_{\scriptscriptstyle X} = \varepsilon \mu_Y + \frac{1}{2} \frac{\sigma_Y^2}{\sigma_X} q_{0,1} \varepsilon^2 + o\left(\varepsilon^2\right).
\end{equation}
 
 Finally, in the case of a Gaussian variable $X$, we can demonstrate that $\alpha_{\scriptscriptstyle X}$ appearing in (\ref{eq:quantile_expansion}) is: 
 \begin{equation}
    \alpha_{\scriptscriptstyle X} = \frac{f'_X(q_X)}{f_X(q_X)} = -\frac{q_X - \mu_X}{\sigma_X^2} = \frac{q_{0,1} }{\sigma_X},
\end{equation}
which confirms that, in the Gaussian case, 
\begin{equation}
    q_{\scriptscriptstyle Z} - q_{\scriptscriptstyle X} = \varepsilon \mu_{\scriptscriptstyle Y} + \frac{\varepsilon^2}{2} \sigma_{\scriptscriptstyle Y}^2 \frac{f'_{\scriptscriptstyle X} \left( q_{\scriptscriptstyle X} \right)}{f_{\scriptscriptstyle X} \left( q_{\scriptscriptstyle X} \right)}  + o\left(\varepsilon^2\right).
\end{equation}

\bibliographystyle{ieeetr}
\bibliography{references}
\balance

\endgroup
\end{document}